\documentclass[10pt]{article}
\usepackage[letterpaper]{geometry}
\usepackage{hicss}
\usepackage{times}
\usepackage[none]{hyphenat}
\usepackage{url}
\usepackage{soul}
\usepackage{latexsym}
\usepackage{indentfirst}
\usepackage{graphicx}
\graphicspath{{images/}}

\usepackage{mathtools}
\usepackage{amsfonts}
\usepackage{tikz}
\usetikzlibrary{decorations.pathreplacing,angles,quotes}
\usepackage{bbm}
\usepackage[hidelinks]{hyperref}
\usetikzlibrary{calc}
\usetikzlibrary{shapes.misc}
\tikzset{cross/.style={cross out, draw=black, minimum size=2*(#1-\pgflinewidth), inner sep=0pt, outer sep=0pt},
cross/.default={1pt}}

\usepackage{pgfplots}    
\usetikzlibrary{calc} 
\usepackage{pgfplots}
\pgfplotsset{compat=1.15}
\usepackage{mathrsfs}
\usetikzlibrary{arrows}
\usepackage{amsmath}
\usepackage{amsfonts}
\usepackage{amssymb}
\usepackage[shortlabels]{enumitem}
\usepackage{blindtext}

\usepackage[ruled,linesnumbered]{algorithm2e}

\usepackage{amsthm}
\usepackage{stmaryrd}

\title{Network Inspection Using Heterogeneous Sensors for Detecting Strategic Attacks} 

\author{Bobak McCann \\
 Georgia Institute of Technology\\
 {\underline{bmccann6@gatech.edu}} \\\And
 Mathieu Dahan \\
 Georgia Institute of Technology \\
 {\underline{mathieu.dahan@isye.gatech.edu} }}

\date{}

\newcommand{\bN}{\ensuremath{\mathbb{N}} }

\newcommand{\tm}[1]{\text{ #1 }}

\newtheorem{theorem}{Theorem}
\newtheorem{corollary}{Corollary}
\newtheorem{lemma}{Lemma}

\newtheorem{proposition}{Proposition}

\theoremstyle{definition}
\newtheorem{example}{Example}

\newlist{Steps}{enumerate}{2}
\setlist[Steps, 1]{leftmargin=1.5cm, label=Step \arabic*. }
\setlist[Steps, 2]{leftmargin=0.7cm, label=\roman*)}

\begin{document}

\maketitle
\begin{abstract}

We consider a two-player network inspection game, in which a defender allocates sensors with potentially heterogeneous detection capabilities in order to detect multiple attacks caused by a strategic attacker. The objective of the defender (resp. attacker) is to minimize (resp. maximize) the expected number of undetected attacks by selecting a potentially randomized inspection (resp. attack) strategy.
We analytically characterize Nash equilibria of this large-scale zero-sum game when every vulnerable network component can be monitored from a unique sensor location. We then leverage our equilibrium analysis to design a heuristic solution approach based on minimum set covers for computing inspection strategies in general. Our computational results on a benchmark cyber-physical distribution network illustrate the performance and computational tractability of our solution approach.

% \bm{Make sure to do a replace all for ``detector"}

% .

% We consider a two-player network inspection game, in which a defender positions heterogeneous sensors according to a probability distribution in order to detect multiple attacks caused by a strategic attacker. We assume the defender has access to multiple types of sensors that can potentially differ in their accuracy. The objective of the defender (resp. attacker) is to minimize (resp. maximize) the expected number of undetected attacks. We derive a class of Nash equilibria of this zero-sum game under the assumption that each component in the network can be monitored  from a unique sensor location. We then leverage our constructed Nash equilibrium to provide approximate solutions to the general case by solving a minimum set cover problem. Our results illustrate the performance and computational advantage of our solution approach, as well as the value of strategically leveraging heterogeneous sensors to protect critical networks against attacks.
\end{abstract}

\section{Introduction}
Critical infrastructure networks such as electric, gas, and water distribution systems are paramount for the well-being of society. However, these networks regularly face random disruptions as well as attacks from strategic adversaries \cite{7011179, inbook1}. In particular, recent incidents have demonstrated that adversarial attackers can disrupt or gain control of the cyber-physical systems deployed in these networks by exploiting cyber insecurities or physical faults.
A most recent example is the cyberattack against a major US fuel pipeline, which caused disruptions in the fuel supply of the Eastern United States
\cite{russon_2021}.
% Another recent example is the power grid failure in Texas, which occurred from an unintentional disruption. 
% \cite{rebecca&nbsp;smith_2021}
Additional examples can be found in \cite{inproceedings1, article3}.

A key part of any defense strategy is to detect attacks using sensors positioned in various locations that continuously monitor the network.
If a network is small, this can be done easily by placing a sensor at each location of interest.
However, for medium or large networks, it can be infeasible to position a sensor at every location.
Thus the problem of how to strategically position a restricted number of sensors is crucial.

We employ a game-theoretic approach to study this problem. 
Game theory has successfully been used to study problems in the domain of cybersecurity (and network security more broadly)
\cite{7011006, MIAO201855, 7498672, 10.1007/978-3-319-47413-7_6, 10.5555/1402795.1402819, article1, BERTSIMAS2016114, DBLP:journals/corr/abs-1903-07261}. In particular, it has proven successful for sensor allocation problems \cite{ DBLP:journals/corr/abs-1903-07261,DBLP:journals/corr/DahanSA17, pirani2021strategic}. 
% \md{Starting from here, the introduction needs to be reorganized. First, you need to describe your model in one paragraph (which you can merge with the previous sentences). Then, you explain how it relates with [14]. Then you show your contributions. The last paragraph is still about how the rest of the paper is structured.}
% \bm{Better? I am a bit confused if what I did is what you're looking for, but if not maybe we can do a bluejeans call.}
In our model, the defender allocates heterogeneous sensors in order to detect multiple attacks caused by a strategic attacker. The sensors may differ in their detection accuracies, which typically depend on the sensing technology utilized. The defender
(resp. the attacker) aims to minimize (resp. maximize) the expected number of undetected attacks.
% Our network model consists of components that an attacker can target simultaneously, and which can be monitored by positioning sensors at predefined corresponding sets of locations (also referred to as nodes).
% At each node, a fixed subset of components can be monitored if a sensor is positioned there.
Thus we model the interactions between both players using a zero-sum game, in which both players may potentially select randomized strategies. This feature is known to be desirable in security settings in which finite resources are allocated
\cite{DBLP:journals/corr/DahanSA17, inproceedings2}.

% Our  game  is  an  extension  of  the  one  in  [20].  The  net-work  consists  of  a  set  of  components  that  can  be  targetsfor  an  attacker,  and  a  set  of  nodes  representing  potentialdetector  locations  from  which  a  subset  of  components  canbe monitored. Nonetheless, while in [20] the allocation of adetector  on  a  node  results  in  a  perfect  detection  of  attackson  its  monitored  components,  we  consider  in  our  modelthat detection capabilities are affected by the locations fromwhich  the  inspection  takes  place.  Namely,  each  node  isassociated with a detection rate that captures the local effectsthat undermine the reliability of a detector

Previous simultaneous security models, such as in \cite{patrolling, garnaev2000search, mavronicolas, Garnaev1997}, assume that each detection device is homogeneous. In this work, we extend the model in \cite{DBLP:journals/corr/DahanSA17} by accounting for the potential heterogeneity in detection accuracy of the sensors available to the defender. In particular, we study how the detection heterogeneity of the defender's resources affects the strategies of both players.

% However, we allow for each sensor to have heterogeneous detection capabilities. 
% This affects the type of solutions that can be obtained.
% \bm{I think we should keep this last sentence in even though we haven't mentioned NE yet because in the reviews they mentioned it should be made clearer whether assuming heterogeneous detection capabilities impacts the solutions.}
% Additionally, for the model in \cite{DBLP:journals/corr/DahanSA17}, it was assumed that the number of attack resources was low.
% However, in our model we do not make this assume this constraint.

We study the mixed Nash Equilibria (NE) of this game.
As this is a zero-sum game, NE can be computed by solving a linear program \cite{article2}.
However, as the network's size increases, this linear program becomes too computationally expensive to solve because of the combinatorial nature of the players' action sets. Thus, we analyze equilibrium properties under certain conditions, and leverage our results to provide a computationally tractable heuristic solution approach that computes inspection strategies in the general case with good detection performance.

Our contributions are twofold: First, we analytically solve the game and provide equilibrium properties when each component in the network is monitored from a unique sensor location. These results provide us with valuable insight regarding the impact of the detection accuracies, number of attacks, and network topology on the players' equilibrium strategies.
Second, we leverage our equilibrium results to design a heuristic solution approach for computing inspection strategies in general. Our approach is based on solutions to a minimum set cover problem, which have been shown to be effective for different inspection games \cite{10.5555/1402795.1402819,article1, BERTSIMAS2016114, DBLP:journals/corr/abs-1903-07261}.
We then conduct a computational study on a benchmark cyber-physical distribution network and empirically validate the performance and computational tractability of our solution approach.

The paper is structured as follows. 
In Section \ref{sec:Problem}, we introduce the network inspection game.
In Section \ref{sec:Disjoint}, we derive equilibrium properties and solve the game when each component is monitored from a unique sensor location. We then present in Section \ref{sec:General} our heuristic approach for computing inspection strategies in the general case and provide computational results to validate our approach. Finally, we summarize our contributions and plans for future work in Section~\ref{sect:pdf}.

\section{Problem Description}\label{sec:Problem}

We consider a network containing a set of vulnerable components $E$ that can be targeted by an attacker. A defender has access to $b_1\in\mathbb{N}$ sensors that can be positioned among 
% \bm{Also, I think we need to mention that $b_1 \leq n$ at some point! -Not really needed since we allow empty sets now}
a set of locations (nodes) $V$ for network monitoring. A sensor positioned at node $v \in V$ monitors a subset of components $E_{v} \subseteq E$, which we refer to as the \emph{monitoring set} of $v$.
% \add{Without loss of generality,} \ch{W}{w}e assume that \add{$(i)$} multiple sensors cannot be simultaneously positioned at the same node 
% \bm{Not sure I agree with this, since the assumption that multiple sensors cannot be simultaneously positioned at the same node isn't really a ``Without loss of generality" thing. I think it should be like ``We assume that multiple sensors cannot be simultaneously positioned at the same node. Additionally, without loss of generality we assume that $(i)$ Each monitoring set nonempty stuff $(ii)$ every component in at least one monitoring set stuff. For ease of exposition, ...."} 
% \ch{.Without loss of generality, we assume that}{, $(ii)$} each monitoring set $E_v$ ($v \in V$) is nonempty, and \add{$(iii)$} \st{that} every component $e \in E$ belongs to at least one monitoring set. 
For ease of exposition, we denote $n \coloneqq |V|$ and $[k] \coloneqq \{1, \dots, k\}$ for every $k\in \bN$.
% \st{We let $[b_1]$ denote our set of sensors.}

We consider that sensors can potentially differ in their detection capabilities.
Specifically, 
for each sensor $k \in [b_1]$, we let $\lambda_k \in (0,1]$ denote its accuracy, i.e., the probability that it detects an attack conducted against a given component within the monitoring set of the node at which it is positioned.
We order the sensors so that $\lambda_1 \geq \dotsm \geq \lambda_{b_1}$.
% \ch{A}{Then, a} \textit{sensor positioning} \ch{is}{can be represented by} \bm{how about instead: ``A \textit{sensor positioning} is then defined to be a vector $s=....$"} a vector $s=(s_1, \dots, s_{b_1}) \in V^{b_1}$ such that $s_i \neq s_j$ for every $(i,j) \in [b_1]^2$ with $i \neq j$, and where $s_k$ represents the node at which sensor $k \in [b_1]$ is positioned by the defender.
Without loss of generality, we assume that multiple sensors cannot be simultaneously positioned at the same node. 
Indeed, positioning additional  sensors at a node $v \in V$ can be equivalently viewed as positioning them among $b_1 - 1$ different copies of node $v$, where each copy has an identical monitoring set $E_v$.
A \textit{sensor positioning} is then represented as a vector $s=(s_1, \dots, s_{b_1}) \in (V\cup \{0\})^{b_1}$ such that $s_i \neq s_j$ for every $(i,j) \in [b_1]^2$ with $i \neq j$ and $s_i, s_j \neq 0$. Here, $s_k \in V$ represents the node at which sensor $k \in [b_1]$ is positioned by the defender, and $s_k = 0$ corresponds to sensor $k$ not being positioned within the network. For consistency, we let $E_0 \coloneqq \emptyset$.
We denote the set of all sensor positionings as $A_1$.

To analyze the problem of strategically positioning sensors in the network, we introduce a zero-sum game $\Gamma \coloneqq \langle \{1,2\}, (\Delta (A_1) , \Delta (A_2) ) , (-U, U) \rangle$. In this game, Player 1 (\textbf{P1}) is the defender who selects a sensor positioning $s \in A_1$. 
Simultaneously, Player 2 (\textbf{P2}) is an attacker who selects a subset of components $T\in 2^E$ to target, where $|T| \leq b_2$ and $b_2\in [|E|]$ is the number of attack resources he has at his disposal.
We refer to such a subset of components as an \textit{attack plan}, and denote the set of all attack plans as $A_2$. 

In such security settings, it may be beneficial for one or both players to randomize their strategies. This feature is especially important for applications where sensing resources can be regularly moved throughout a network, which increases the strategic uncertainty faced by the attacker and hence generally achieves a higher protection level \cite{10.5555/1402795.1402819,Hochbaum2011}.
Thus, we allow \textbf{P1} and \textbf{P2} to select mixed strategies.
A \emph{mixed strategy} for the defender (resp. attacker) is a probability distribution over the set of sensor positionings $A_1$ (resp. the set of attack plans $A_2$).
Namely, we define the set of mixed inspection and attack strategies as 
$\Delta(A_1) \coloneqq \{\sigma^1 \in [0,1]^{|A_1|} \mid \sum_{s  \in A_1} \sigma^1_{s}  = 1 \}$ and $\Delta(A_2) \coloneqq \{\sigma^2 \in [0,1]^{|A_2|} \mid \sum_{T  \in A_2} \sigma^2_T  = 1 \}$ respectively, where $\sigma_{s}^1$ (resp. $\sigma_T^2$) represents the probability assigned to the sensor positioning $s \in A_1$ (resp. the attack plan $T \in A_2$) under the inspection strategy $\sigma^1$ (resp. the attack strategy $\sigma^2$). 
% \md{I don't like the following sentence. I don't believe it adds new information. We may/should have one sentence on how we translate $\sigma$s into scheduling of operations.}
% \bm{Agreed}
% Intuitively, we can think of \textbf{P1} (resp. \textbf{P2}) as choosing which
% sensor positioning she implements (resp. which attack plan he implements) according to the probability distribution $\sigma^1$ (resp. $\sigma^2$). 
We assume that the players' strategies are independent randomizations.

In this model, we assume that the sensors are safe from possible damage during an attack; only the components in the network can be targeted. 
Additionally, we assume that detection is independent across attacks and sensors, and that if an attack against a component is detected, then the defender can nullify the damage.
Hence, in our model we consider an attack on a component by \textbf{P2} to be successful if and only if it is not detected by \textbf{P1}. 
% \st{We also assume that the probabilities that positioned sensors detect a same attack within their monitoring sets are independent of one another.}
As such, \textbf{P1} (resp. \textbf{P2}) seeks to minimize (resp. maximize) the expected number of undetected attacks which, for any strategy profile $(\sigma^{1}, \sigma^{2}) \in \Delta(A_1) \times \Delta(A_2)$, is given by
$$ U(\sigma^1, \sigma^2) \coloneqq \mathbb{E}_{(\sigma^1, \sigma^2)} \left[ \sum_{e \in T} \prod_{k=1}^{b_1} \left(1-\lambda_k \mathbbm{1}_{\{e\in E_{s_k}\}} \right) \right],$$ 
% Here, we are calculating the expected number of undetected attacks over all pairs $(s, T)\in A_1\times A_2$, where the probability assigned to $(s, T)$ in the expectation is given by $\sigma^1_s \cdot\sigma^2_T$. \bm{Bad sentence I just added? Do I need to rephrase?}\md{Yes indeed.}\add{[I have changed the period at the end of the equation by a comma] where the expectation is taken over all pairs of actions $(s, T)\in A_1\times A_2$ that are selected with probability $\sigma^1_s \cdot\sigma^2_T$ by the players' strategies/the strategy profile $(\sigma^1,\sigma^2)$. [something like this]} \bm{to discuss in meeting}
% % \add{In particular, this expression calculates, for each targeted component, the probability that this attack remains 
where the expectation is taken over all pairs of actions $(s, T)\in A_1 \times A_2$, which are selected with probability $\sigma^1_s\cdot \sigma^2_T$ by the players' strategies.
% Additionally, if $s_k = 0$ for some $k \in [b_1]$, i.e., sensor $k$ is not positioned in the network, then we let $E_{0} \coloneqq \emptyset$.
% \bm{I think should be rewritten to be:
% ``
% where the expectation is taken over all pairs of actions $(s, T)\in A_1 \times A_2$, which are selected with probability $\sigma^1_s\cdot \sigma^2_T$ by the players' strategies, and where we let $E_{\emptyset} \coloneqq \emptyset$.
% "}
% {In this expression, we compute, for every pair of actions $(s,T) \in A_1\times A_2$ selected with probability $\sigma^1_s \cdot \sigma^2_T$ by both players, the sum over each }

Next, we show an instantiation of the zero-sum game $\Gamma$ via an example.

\begin{example}
We consider an example of a network represented in Figure~\ref{fig:detection_model}.

\begin{figure}[htbp]
    \centering
        \begin{tikzpicture}[scale=1.25,main node/.style={circle,draw,inner sep = 0.15cm},main node2/.style={circle,draw,inner sep = 0.10cm},]
            \fill[fill=gray!30!white] (-0.3, 0) ellipse [x radius = 15mm, y radius = 6mm];
            \fill[fill=gray!30!white] (1.5, 0) ellipse [x radius = 15mm, y radius = 5mm];
            \fill[fill=gray!30!white] (0.5, -1) ellipse [x radius = 5mm, y radius = 20mm];
            \fill[fill=gray!30!white] (1.5, -2) ellipse [x radius = 20mm, y radius = 6mm];
            \fill[fill=gray!30!white] (2, -2) ellipse [x radius = 5mm, y radius = 5mm];
            
            \draw (-0.3, 0) ellipse [x radius = 15mm, y radius = 6mm];
            \draw (1.5, 0) ellipse [x radius = 15mm, y radius = 5mm];
            \draw (0.5, -1) ellipse [x radius = 5mm, y radius = 20mm];
            \draw (1.5, -2) ellipse [x radius = 20mm, y radius = 6mm];
            \draw (2, -2) ellipse [x radius = 5mm, y radius = 5mm];
            
            % \filldraw[fill=gray!50!white] (0, 0) ellipse [x radius = 15mm, y radius = 5mm];
            % \filldraw[fill=gray!50!white] (1.5, 0) ellipse [x radius = 15mm, y radius = 5mm];
            % \filldraw[fill=gray!50!white] (0.5, -1) ellipse [x radius = 5mm, y radius = 20mm];
            % \filldraw[fill=gray!50!white] (1.5, -2) ellipse [x radius = 20mm, y radius = 6mm];
            % \filldraw[fill=gray!50!white] (2, -2) ellipse [x radius = 5mm, y radius = 5mm];
            % \filldraw[fill=blue] (-1, 0) circle [radius=0.25];
            
            \node (1) at (-1.5, 0) {$v_1$};
            \node (2) at (0.5, 0.68) {$v_2$};
            \node (3) at (2.75, 0) {$v_3$};
            \node (4) at (-0.2, -2.1) {$v_4$};
            \node (5) at (1.9, -2.3) {$v_5$};
            
            \node[circle, draw, fill=blue!50!white, scale=0.7] (e_1) at (-0.95, 0) {$e_1$};
            \node[circle, draw, fill=blue!50!white, scale=0.7] (e_2) at (-0.3, 0) {$e_2$};
            \node[circle, draw, fill=blue!50!white, scale=0.7] (e_3) at (0.5, 0) {$e_3$};
            \node[circle, draw, fill=blue!50!white, scale=0.7] (e_4) at (1.5, 0) {$e_4$};
            \node[circle, draw, fill=blue!50!white, scale=0.7] (e_5) at (2.15, 0) {$e_5$};
            \node[circle, draw, fill=blue!50!white, scale=0.7] (e_6) at (0.5, -1) {$e_6$};
            \node[circle, draw, fill=blue!50!white, scale=0.7] (e_7) at (0.5, -2) {$e_{7}$};
            \node[circle, draw, fill=blue!50!white, scale=0.7] (e_8) at (2.05, -1.8) {$e_{8}$};
            \node[circle, draw, fill=blue!50!white, scale=0.7] (e_9) at (3, -2) {$e_{9}$};
            
            \node (green sensor) at (-0.9, -1.1) {\includegraphics[scale=0.1]{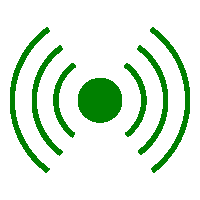}};
            \node (yellow sensor) at (1.8, 0.95) {\includegraphics[scale=0.1]{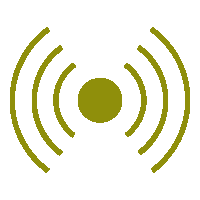}};            
            \draw[->, thick, blue] (green sensor) edge[bend left] node [left] {$s'$} (1);
            \draw[->, thick, red] (green sensor) edge[bend left] node [left] {$s$} (4);
            \draw[->, thick, blue] (yellow sensor) edge[bend right] node [left] {} (0.65, 0.72);
            \node[blue] at (1, 1.2) {$s'$};
            \draw[->, thick, red] (yellow sensor) edge[bend left] node [left] {$s$} (3);
            
            \draw (e_1) node[cross=7pt, orange!80!black, thick,opacity=0.8] {};
            \node[orange!80!black] at (-0.8, 0.3) {\footnotesize{$T$}};
            \draw (e_3) node[cross=7pt,  orange!80!black, thick,opacity=0.8] {};
            \node[orange!80!black] at (0.66, 0.28) {\footnotesize{$T$}};
            
            \draw (e_6) node[cross=7pt, red!70!black, thick,opacity=0.8] {};
            \node[red!50!black] at (0.78, -0.73) {\footnotesize{$T'$}};

        \end{tikzpicture}
    \caption{Game instance on a network containing 5 nodes and 9 components.} 
    \label{fig:detection_model}
\end{figure}
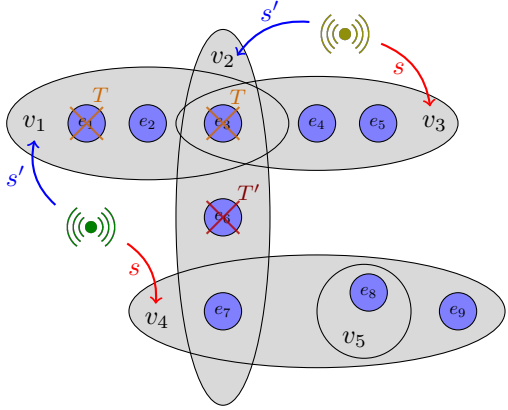

In this example, the set of nodes is $V = \{v_1, \dots, v_5\}$, the set of components is $E = \{e_1, \dots, e_9\}$, and the monitoring sets are $E_{v_1} = \{e_1, e_2, e_3\},$ $E_{v_2} = \{e_3, e_6, e_7\}, E_{v_3} = \{e_3, e_4, e_5\}, E_{v_4} = \{e_7, e_8, e_9\}$, and $E_{v_5} = \{e_8\}$. The defender has two sensors. Sensor 1 (in green) has accuracy $\lambda_1 = 0.9$, and sensor 2 (in yellow) has accuracy $\lambda_2 = 0.5$. In this example, the defender selects the randomized inspection strategy $\sigma^1$ defined by $\sigma^1_s = 0.4$ and $\sigma^1_{s'} = 0.6$, with $s = (v_4, v_3)$ and $s' = (v_1, v_2)$. Simultaneously, the attacker selects the randomized attack strategy $\sigma^2$ defined by $\sigma^2_{T} = 0.2$ and $\sigma^2_{T'} = 0.8$, with $T = \{e_1, e_3\}$ and $T' = \{e_4\}$.
For this example, the expected number of undetected attacks is given by
\begin{align*}
    U(\sigma^1, \sigma^2) &=\hspace{0.1cm} \sigma^1_s \sigma^2_T\left( 1 + (1 - \lambda_2)  \right) \\
     &\hspace{0.15cm}+ \sigma^1_s \sigma^2_{T'} (1) \\
     &\hspace{0.15cm}+ \sigma^1_{s'} \sigma^2_{T} \left((1 - \lambda_1) + (1-\lambda_1)(1-\lambda_2) \right) \\
    &\hspace{0.15cm}+ \sigma^1_{s'}\sigma^2_{T'}(1-\lambda_2)\\
    &= 0.698.
\end{align*}
\hfill$\triangle$
\end{example}

In simultaneous games, a solution concept is given by Nash Equilibrium. 
Specifically, a strategy profile 
$(\sigma^{1*}, \sigma^{2*}) \in \Delta(A_1) \times \Delta(A_2)$ is a \emph{Nash Equilibrium} (NE) of $\Gamma$ if for all $(\sigma^1, \sigma^2) \in \Delta(A_1)   \times \Delta(A_2)$, we have
\begin{align*}
    U(\sigma^{1*}, \sigma^2)  \leq U(\sigma^{1*}, \sigma^{2*}) \leq U(\sigma^1, \sigma^{2*}).
\end{align*}

Equivalently, at a NE, $\sigma^{1*}$ (resp. $\sigma^{2*}$) is a best response to $\sigma^{2*}$ (resp. $\sigma^{1*}$).
% \bm{``Equivalently, $(\sigma^{1*}, \sigma^{2*})$ is a NE if $\sigma^{1*}$ (resp. $\sigma^{2*}$) is a best response to $\sigma^{2*}$ (resp. $\sigma^{1*}$)."}
We refer to $\sigma^{1*}$ (resp. $\sigma^{2*}$) as an equilibrium inspection strategy (resp. equilibrium attack strategy). 
Additionally, we refer to $U(\sigma^{1*}, \sigma^{2*})$ as the \emph{value of the game}. 
% Since $\Gamma$ is a finite zero-sum game, the value $U(\sigma^{1*}, \sigma^{2*})$  is \add{ defined and} identical for every strategy profile $(\sigma^{1*}, \sigma^{2*})\in \Delta(A_1)\times \Delta(A_2)$ that is a NE. In other words, the value of the game is unique and well-defined.
Since $\Gamma$ is a finite zero-sum game, the value $U(\sigma^{1*}, \sigma^{2*})$ exists and is identical for every strategy profile $(\sigma^{1*}, \sigma^{2*})\in \Delta(A_1)\times \Delta(A_2)$ that is a NE. In other words, the value of the game is unique and well-defined.

% Furthermore, the set of NE and the value of the zero-sum game $\Gamma$ can be obtained by solving the following pair of dual linear programming (LP) problems \cite{article2}:
% \begin{align*}
%     &(P)  \min_{\sigma^1 \in \Delta(A_1)} \max_{T\in A_2}U(\sigma^1, T)\\
%     &(LP_2) \max_{\sigma^2 \in \Delta(A_2)} \min_{s \in A_1}U(s , \sigma^2).
% \end{align*}
 
%  \md{In the latex file, the text above what I'm writing here was not commented out. Please make sure I didn't make any mistake by commenting it out.}
 
Furthermore, the zero-sum game $\Gamma$ can be solved using the following linear programming problem \cite{article2}:
\begin{align*}
    (\mathcal{P}) \  \min_{\sigma^1 \in \Delta(A_1)} \max_{T\in A_2}U(\sigma^1, T).
\end{align*}
Specifically, the equilibrium inspection strategies, equilibrium attack strategies, and value of the game $\Gamma$ are given by the optimal primal solutions, optimal dual solutions, and optimal value of $(\mathcal{P})$, respectively. 
 
%  \md{We never use $(LP_2)$ in this paper. I propose the following change. This would mean changing every occurrence of $(LP_1)$: Furthermore, the zero-sum game $\Gamma$ can be solved using the following linear programming (LP) problem \cite{article2}:
% \begin{align*}
%     (\mathcal{P}) \  \min_{\sigma^1 \in \Delta(A_1)} \max_{T\in A_2}U(\sigma^1, T).
% \end{align*}
% Specifically, the equilibrium inspection strategies, equilibrium attack strategies, and value of the game $\Gamma$ are given by the optimal primal solutions, optimal dual solutions, and optimal value of $(\mathcal{P})$, respectively.
% }

However, solving $(\mathcal{P})$ becomes intractable even for medium-sized networks due to the combinatorial nature of the players' sets of actions: the number of variables and constraints in $(\mathcal{P})$ are given by $1+|A_1|  = 1+\sum_{i=0}^{b_1}i!\binom{n}{i}$ and $1+|A_2| = 1+\sum_{j=0}^{b_2} \binom{|E|}{j}$, respectively.
Thus, in this paper, we present an approach to provide approximate solutions to the game $\Gamma$. We first derive an analytical characterization of a class of NE when the monitoring sets are mutually disjoint. We then leverage this result in Section~\ref{sec:General} to derive a heuristic method for computing an approximate solution in general.

% Finally, we note that in a zero-sum game, no player has a first-mover advantage. This implies that if the players were to play sequentially, the \ch{NE}{equilibrium solutions} would remain unchanged[/ or valid?]. 
% Thus, \ch{our}{the} game \add{$\Gamma$} can be used to model the scenario when the attacker chooses his attack strategy after observing \ch{what the defender chose as her inspection strategy}{the defender's inspection strategy}. This type of situation is frequently encountered in cybersecurity applications, and various other \add{security} applications more generally.

Henceforth, we assume without loss of generality that $(i)$ $b_1\leq n$, $(ii)$ $b_2 \leq |E|$, $(iii)$ each monitoring set $E_v \; (v\in V)$ is nonempty, and $(iv)$ every component $e\in E$ belongs to at least one monitoring set. Indeed, if some components do not belong to any monitoring set, then \textbf{P2} will always target these components and allocate his remaining resources among the components that belong to at least one monitoring set.

Our game models scenarios where, for instance, each component represents an asset that can be hacked, and each node represents a computer on which software protocols can be installed to detect cyber attacks. In this scenario, our sensors are the software security protocols, which each have a certain probability of detecting a cyber attack. Stronger protocols are harder to be bypassed, and will detect an intrusion with a higher probability than a weaker protocol, which a hacker can more easily bypass.

Finally, we note that in a zero-sum game, no player has a first-mover advantage. This implies that if the players were to play sequentially, the equilibrium solutions would remain valid. Thus, the game $\Gamma$ can be used to model scenarios where the attacker selects his attack strategy after observing the defender's inspection strategy. This type of situation is frequently encountered in cybersecurity applications and various other security problems more broadly.

\section{Game-Theoretic Analysis for Mutually Disjoint  Monitoring Sets}\label{sec:Disjoint}

% \bm{Make sure pronouns are consistent. Later I switched to using he instead of she for the defender}
% \bm{Also another general note: Make sure in the end to do a search-and-replace for ``detector" with ``sensor" just in case I accidentally wrote ``detector" somewhere instead of ``sensor".}

In this section, we study the game $\Gamma$ when all the monitoring sets are mutually disjoint. That is, when $E_{v}\cap E_{w} = \emptyset$  for all $(v, w)\in V^2$ such that $v\neq w$.
Without loss of generality, we rewrite the set of nodes as $V = \{v_1,\dots,v_n\}$ so that $|E_{v_1}| \geq \dotsm \geq |E_{v_n}|$. 
Furthermore, to simplify the equilibrium analysis, we define for every $(\sigma^1, v)\in \Delta(A_1) \times V$
the \textit{detection probability} of node $v$ under $\sigma^1$ as:
% inspection strategy $\sigma^1 \in \Delta(A_1)$ and node $v \in V$ the resulting \emph{detection probability} of $v$ as follows:
$$ p_{\sigma^1}(v)\coloneqq \sum\limits_{j=1}^{b_1} \lambda_j \sum\limits_{\quad\quad\mathclap{\{s \in A_1 \mid s_j = v\}}\quad\quad} \sigma^1_s.$$
That is, $p_{\sigma^1}(v)$ represents the probability that an attack in the monitoring set $E_v$ is detected under the inspection strategy $\sigma^1$.

Similarly, we define for every $(\sigma^2, e)\in \Delta(A_2)\times E$
the \textit{attack probability} of component $e$ under $\sigma^2$ as
$$p_{\sigma^{2}}(e) \coloneqq \sum_{\quad\quad\mathclap{\{ T \in A_2 \mid e \in T \}}\quad\quad}  \sigma^2_T.$$
% \bm{How about instead ``Similarly, for every $(\sigma^2, e)\in \Delta(A_2)\times E$ we define the \textit{attack probability of $e$ under $\sigma^2$ as 
% $$p_{\sigma^{2}}(e) \coloneqq \sum_{\{ T \in A_2 \mid e \in T \}}  \sigma^2_T.$$}}
That is, $p_{\sigma^2}(e)$ represents the probability that $e$ is targeted under the attack strategy $\sigma^2$.

In order to maximize the expected number of undetected attacks, \textbf{P2}'s incentive is to spread his attacks across the monitoring sets, thus making it more challenging for \textbf{P1} to detect the attacks. However, \textbf{P2} is constrained by the topology of the network, and more particularly by the sizes of the different monitoring sets. This in turn will impact \textbf{P1}'s best-response inspection strategy.

More formally, we consider the following quantity:
$$k^* = \min \left\{ k\in [n] \, \middle| \, \frac{b_2 - \sum_{j=k+1}^n |E_{v_j}|}{k} \geq |E_{v_{k+1}}| \right\},$$
where we let $|E_{v_{n+1}}| \coloneqq 0$.
% \bm{Would ``convention" be the right word here? I would think ``convention" would be more for things that a lot of people use and for something has been around for a while. Would the word ``assumption" would be better?}. 
% \md{We need a one sentence high-level explanation of $k^*$ here.}
% \bm{How about: ``Essentially, $k^*$  represents the size of the smallest subset of nodes $S$ where we are guaranteed that, on average, each node in $S$ will have more components in its monitoring set targeted than the number of components targeted in the monitoring set of any node outside $S$" ?}
% \bm{the monitoring sets that are not fully targeted.}
Essentially, $\{E_{v_1}, \dots, E_{v_{k^*}} \}$ represents the monitoring sets that are not fully targeted by \textbf{P2} when he spreads his attacks.
% \bm{How about we get rid of ``once" and replace it with ``when"?}

The next theorem then characterizes a class of NE of the game $\Gamma$ when the monitoring sets are mutually disjoint: 
\begin{theorem}
\label{thm:NE}
    If $E_{v}\cap E_{w} = \emptyset$  for all $(v, w)\in V^2$ such that $v\neq w$, then a strategy profile $(\sigma^{1*}, \sigma^{2*}) \in \Delta(A_1) \times \Delta(A_2)$ is a NE if it satisfies the following conditions: 
    % {\scriptsize\bm{Change this away from $i$ because we use that for nodes} \bm{-I did this in other parts, but here I think I should keep it as $i$ since the lambdas are tied to the nodes}}
    % \begin{align}
    %   {p_{\sigma^{1*}}(v_i)}  &=\hspace{-0.05cm}
    %     \begin{cases}
    %         \dfrac{\sum\limits_{j=1}^{k^*}\lambda_j}{k^*} & \text{if } 1 \leq i \leq k^*, \\
    %         \lambda_i & \text{if } k^* < i \leq b_1, \\
    %         0 & \text{if } b_1 < i \leq  n,
    %     \end{cases}\label{NE:detection_prob}\\[0.2cm]
    %     \sum\limits_{e\in E_{v_i}} \hspace{-0.08cm}p_{\sigma^{2*}}(e) &= \hspace{-0.05cm}
    %     \begin{cases}
    %         \dfrac{b_2 - \,\sum\limits_{\mathclap{j=k^*+1}}^n \,|E_{v_j}|}{k^*} & \text{if }  1 \leq i\leq k^*,\\
    %         |E_{v_i}|  &   \text{if } k^*<i \leq n.
    %     \end{cases}\label{NE:attack_prob}
    % \end{align}
% 
% \md{Now, I believe there is an issue with (1). That formula works when $k^* \leq b_1$. If $k^* > b_1$, then (1) is ill-defined: Sure, the second case is empty, but the first and third cases overlap. If this is indeed the case, I will be honest: I'm not happy. This is a big mistake that would be rather bad if the paper were to be published this way.}
% 
% If $E_{v}\cap E_{w} = \emptyset$  for all $(v, w)\in V^2$ such that $v\neq w$, then the following properties hold:
% 
% \begin{itemize}[leftmargin=*]
%     \item[--] An inspection strategy $\sigma^{1*} \in \Delta(\mathcal{A}_1)$ is an equilibrium inspection strategy if it satisfies the following conditions:
    \begin{align}
      {p_{\sigma^{1*}}(v_i)}  =\hspace{-0.05cm}
        \begin{cases}
            \dfrac{1}{k^*}\quad \sum\limits_{j=1}^{\mathclap{\min\{b_1,k^*\}}}\, \lambda_j & \text{if } 1 \leq i \leq k^* \\
            \lambda_i & \text{if } k^* < i \leq b_1 \\
            0 & \text{if } \max\{b_1,k^*\} < i \leq  n,
        \end{cases}\label{NE:detection_prob}\\
         \,\sum_{\mathclap{e\in E_{v_i}}} p_{\sigma^{2*}}(e) = \hspace{-0.05cm}
        \begin{cases}
            \dfrac{1}{k^*}\left( b_2 - \,\sum\limits_{\mathclap{j=k^*+1}}^n \,|E_{v_j}| \right)\hspace{-0.05cm} & \text{if }  1 \leq i\leq k^*\\
            |E_{v_i}|  &   \text{if } k^*<i \leq n.
        \end{cases}\label{NE:attack_prob}
    \end{align}
    
    % \bm{I see that in the pdf there is a (1) for the attacker strategy condition, but I don't see where that is in the latex?}
    
  Furthermore, the value of the game is given by the following expression:
    $$
    b_2 - \sum_{i=1}^{b_1} \lambda_i \min\left\{|E_{v_i}| \, , \;\, \dfrac{1}{k^*} \left( b_2 - \sum\limits_{\mathclap{j=k^*+1}}^{n}|E_{v_j}| \right)\right\}.   
    $$ 
% \end{itemize}

% \md{Here is what I believe is one option, but I don't like it 
% \begin{align*}
%   {p_{\sigma^{1*}}(v_i)}  &=\hspace{-0.05cm}
%     \begin{cases}
%         \dfrac{1}{k^*}\sum\limits_{j=1}^{\min\{b_1,k^*\}}\lambda_j & \text{if } 1 \leq i \leq k^*, \\
%         \lambda_i & \text{if } k^* < i \leq b_1, \\
%         0 & \text{if } \max\{b_1,k^*\} < i \leq  n,
%     \end{cases}
% \end{align*}

% }

    % Furthermore, the value of the game is given by
    % $$
    % b_2 - \dfrac{b_2 - \,\sum\limits_{\mathclap{j=k^*+1}}^n \,|E_{v_j}|}{k^*} \cdot \sum\limits_{i=1}^{\min\{b_1, k^*\}} \lambda_i - \sum\limits_{i = k^*+1}^{b_1} \lambda_i |E_{v_i}|.    
    % $$    

\end{theorem}

From Theorem~\ref{thm:NE}, we obtain that when the monitoring sets are mutually disjoint, a class of NE can be  described analytically using the players' resources and the sizes of the monitoring sets. In particular, we find that in equilibrium, \textbf{P2} targets all the components in $E_{v_{k^*+1}},\dots,E_{v_n}$, and allocates his remaining resources uniformly among the first $k^*$ monitoring sets $E_{v_1},\dots,E_{v_{k^*}}$. By definition of $k^*$, we have that: 
$$\frac{b_2 - \sum_{j=k^*+1}^n |E_{v_j}|}{k^*} \geq \left| E_{v_{k^*+1}} \right|.$$

Therefore, since \textbf{P1} aims to minimize the number of undetected attacks, her incentive, given \textbf{P2}'s equilibrium attack strategy, is to position her best sensors (i.e., those with the highest accuracy) among the nodes $\{v_1,\dots,v_{k^*}\}$. 
Moreover, since \textbf{P2} targets all components in $E_{v_{k^*+1}},\dots,E_{v_n}$, \textbf{P1}'s incentive is to position her next best sensor (if available) to the remaining node with the largest monitoring set, namely $v_{k^*+1}$. \textbf{P1} then repeats this process until all her sensors are positioned.

Since the monitoring sets $E_{v_1},\dots,E_{v_{k^*}}$ are not fully targeted under \textbf{P2}'s equilibrium attack strategy, \textbf{P1} must randomize the positioning of her best sensors among the nodes $\{v_1,\dots,v_{k^*}\}$ to ensure that \textbf{P2} does not have an incentive to deviate from his strategy. Thus, \textbf{P1}'s equilibrium inspection strategy is such that the detection probability of each node in $\{v_1,\dots,v_{k^*}\}$ is identical, and given by $\frac{1}{k^*}\sum_{j=1}^{\min\{b_1,k^*\}}\lambda_j$.
In the next lemma, we construct a strategy profile that satisfies the detection and attack probability conditions \eqref{NE:detection_prob}-\eqref{NE:attack_prob} of Theorem~\ref{thm:NE}:
% In the next Lemma, we construct an inspection strategy that satisfies the detection probability condition \eqref{NE:detection_prob} of Theorem~\ref{thm:NE}:

% \bm{Also make sure to have periods in the right places etc.}

\begin{lemma}
\label{lemma:cycling}
\ 
\begin{enumerate}[leftmargin=*]
    \item 

    \begin{itemize}[leftmargin=*]
        \item[--] If $b_1 \leq k^*$, consider for every $l \in [k^*]$ the following sensor positioning:
        % \bm{Do we need to mention that its for each $\l\in [k^*]$ here?}
        % \bm{Note to self: Remains to be checked, but after other stuff since the text spills over into the other column right now}:
        \begin{align*}
            s^l \coloneqq \begin{cases}
                 (v_l,\dots,v_{l+b_1-1})\\
                 \hspace{2cm}\text{if } 1 \leq l\leq k^* - b_1 + 1\\
                 (v_l,\dots,v_{k^*},v_1,\dots,v_{l+b_1 - k^*-1})\\
                 \hspace{2cm}\text{if } k^* - b_1 + 1 < l \leq k^*.
            \end{cases}
        \end{align*}
        
        \item[--] If $b_1 > k^*$, consider for every $l \in [k^*]$ the following sensor positioning:
        \begin{align*}
            s^l \coloneqq \begin{cases}
                 (v_1,\dots,v_{k^*},v_{k^*+1},\dots,v_{b_1})\\ \hspace{4cm}\text{if } l=1\\
                 (v_l,\dots,v_{k^*},v_1,\dots,v_{l-1},v_{k^*+1},\dots,v_{b_1}) \\
                 \hspace{4cm}\text{if } 1 < l \leq k^*.
            \end{cases}
        \end{align*}
   
    \end{itemize}
    
    Then, $\sigma^{1*} \in \Delta(A_1)$ defined by 
    \begin{align*}
        &\sigma^{1*}_{s^l} = \frac{1}{k^*} \ \forall l \in [k^*], \text{ and } \sigma^{1*}_s = 0 \text{ otherwise,}
    \end{align*}
    satisfies condition (\ref{NE:detection_prob}) in Theorem \ref{thm:NE}.
    
    \item Let $b_2^\prime \coloneqq k^* \left\lfloor
    \frac{1}{k^*}\left(b_2 - \sum_{j=k^*+1}^n |E_{v_j}|\right) \right\rfloor$, and for $l\in [k^*]$ let
    \begin{align*}
        &C^l \coloneqq \{ 1, \dots, l+b_2 - b_2^\prime - k^*-1 \} \\
        &\hspace{1cm}\cup \{ l , \dots, \min\{l+b_2 - b_2^\prime-1,k^*\} \}.
    \end{align*}
% \md{I'm not sure I understand your comment. Are you saying you don't want to use that definition of $C^l$: $C^l \coloneqq \{ 1, \dots, l+b_2 - b_2^\prime - k^*-1 \} \cup \{ l , \dots, \min\{l+b_2 - b_2^\prime-1,k^*\} \}$ }
% \bm{Implement these changes he suggested, and have the $C^l$ in-line}

    % \bm{The idea behind the $C$ is that this is the set you are going to cycle over. I think it is cleaner than if I just wrote it in its full form in the piecewise definitions}
    % \bm{Also, note that $b_2'$ is the number of attack resources that are used to fill up the first $|E_{v_k^* + 1}|$ layers of the monitoring sets $1, \dots, k^*$. Thus $b_2 - b_2'$ is the number of attacks that are in the last layer that will be cycled.}
    % \bm{Also, note that if $b_2 - b_2' = 0$, then from the definition we would get $C = \{l, \dots, l-1\} = \emptyset$, so this definition of $C$ still works in that case.}

    Consider attack plans $T^l \; (l\in [k^*])$ defined as follows:
    \begin{align*}
        \hspace{-1cm}\left| T^l \cap E_{v_j} \right|
        \coloneqq
        \begin{cases}
            \frac{b_2^\prime}{k^*}  + 1  & \text{if } j\in C^l\\[5pt]
            \frac{b_2^\prime}{k^*}  & \text{if }  j\in [k^*] \setminus C^l\\
            \left| E_{v_j} \right| &\text{if } k^* < j \leq n.
        \end{cases}
    \end{align*}

    Then, $\sigma^{2*} \in \Delta(A_2)$ 
    defined by
    % \begin{align*}
    %     \sigma^{1*}_{s}
    %     =
    %     \begin{cases}
    %         \frac{1}{k^*} & \text{if there exists } l\in [k^*] \text{ with } s=s^l\\
    %         0             & \text{otherwise}
    %     \end{cases}     
    % \end{align*}
    \begin{align*}
    &\sigma^{2*}_{T^l} = \frac{1}{k^*} \ \forall l \in [k^*], \text{ and } \sigma^{2*}_T = 0 \text{ otherwise,}
    \end{align*}
    % satisfies condition (\ref{NE:detection_prob}) in Theorem~\ref{thm:NE}.
    satisfies condition (\ref{NE:attack_prob}) in Theorem~\ref{thm:NE}.

    \end{enumerate}

\end{lemma}

From Lemma~\ref{lemma:cycling}, we find that an equilibrium inspection strategy can be constructed by ``cycling'' the positioning of sensors  $1, \dots, \min\{k^*,b_1\}$ among the nodes $v_1, \dots, v_{k^*}$: $s^1$ positions sensor 1 at node $v_1$, sensor 2 at node $v_2$, and so on. Then, $s^2$ positions sensor 1 at node $v_2$, sensor 2 at node $v_3$ and so on. Furthermore, if $b_1 > k^*$, then \textbf{P1} deterministically positions sensors $k^* + 1, \dots, b_1$ at the remaining nodes, in decreasing order of their monitoring sets' size: she positions sensor $k^*+1$ at node $v_{k^*+1}$, sensor $k^*+2$ at node $v_{k^*+2}$, and so on.

% \bm{If we change the Lemma to have attack strategies added back, then mention a paragraph about the equilibrium attack strategies like below commented paragraph}
Similarly, an equilibrium attack strategy can be constructed by first deterministically targeting all the components in $E_{v_{k^*+1}},\dots,E_{v_n}$. Then, $\left\lfloor \frac{1}{k^*} \left(b_2 - \sum_{i=k^*+1}^{n} | E_{v_i} |  \right) \right\rfloor$ components are deterministically targeted within each monitoring set in $E_{v_1},\dots,E_{v_{k^*}}$. Finally, \textbf{P2} ``cycles'' his remaining attack resources (if any are remaining) over the remaining components in $E_{v_1},\dots,E_{v_{k^*}}$.

Next, we illustrate Theorem~\ref{thm:NE} and Lemma ~\ref{lemma:cycling} with an example.

\begin{example} 
Consider the network shown in Figure~\ref{fig:example_net}.
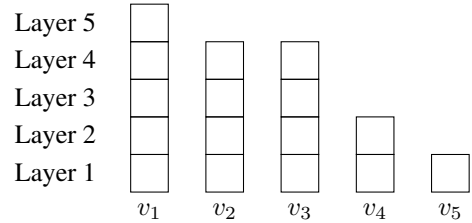
\begin{figure}[htbp]
    \centering
    \begin{tikzpicture}
    \foreach \y in {0, ..., 4}
    {
        \draw (0, 0.5*\y) rectangle (0.5 , 0.5 + 0.5*\y);
    }
    \node at (0.25 , -0.25) {$v_1$};
    
    \foreach \y in {0, ..., 3}
    {
        \draw (0 + 1, 0.5*\y) rectangle (0.5 + 1, 0.5 + 0.5*\y);
    }
    \node at (0.25 + 1 , -0.25) {$v_2$};
    
    \foreach \y in {0, ..., 3}
    {
        \draw (0 + 2, 0.5*\y) rectangle (0.5 + 2, 0.5 + 0.5*\y);
    }
    \node at (0.25 + 2, -0.25) {$v_3$};    
    
    \foreach \y in {0, ..., 1}
    {
        \draw (0 + 3, 0.5*\y) rectangle (0.5 + 3, 0.5 + 0.5*\y);
    }
    \node at (0.25 + 3, -0.25) {$v_4$};
    
    \foreach \y in {0, ..., 0}
    {
        \draw (0 + 4, 0.5*\y) rectangle (0.5 + 4, 0.5 + 0.5*\y);
    }
    \node at (0.25 + 4, -0.25) {$v_5$};    
    
    % \foreach \y in {0, ..., 1}
    % {
    %     \draw (0 + 5, 0.5*\y) rectangle (0.5 + 5, 0.5 + 0.5*\y);
    % }
    % \node at (0.25 + 5, -0.25) {$v_6$};  
    
    \foreach \y in {1, ..., 5}
    {
        \node at (-1, 0.23 + 0.5*\y - 0.5) {Layer \y};    
    }
    
\end{tikzpicture}
    \caption{Example of a network with 5 nodes, 16 components, and mutually disjoint monitoring sets.}
    \label{fig:example_net}
\end{figure}

In this illustration, each square represents a component that can only be monitored from the node indicated below it. Thus, in this example, $|E_{v_1}| = 5$, $|E_{v_2}| = 4$, $|E_{v_3}| = 4$, $|E_{v_4}| = 2$, and $|E_{v_5}| = 1$.
To simplify our equilibrium description, let  $e_{i, j} \; (i\in [n], \; j\in [|E_{v_i}|])$ represent the component in layer $j$ of monitoring set $E_{v_i}$.
This example can be used to represent a computer network in which each computer lies within a closed section of the network, and such that each computer in a given closed section can detect cyberattacks conducted against only the components in its section.

Suppose that \textbf{P1} has 4 sensors. Furthermore, we consider that \textbf{P2} has $b_2 = 10$ attack resources. To spread his attacks in equilibrium, \textbf{P2} can first allocate 5 attack resources to target one component in each monitoring set (in layer 1). Then \textbf{P2} can allocate 4 attack resources to target one more component in each monitoring set that is not fully targeted (in layer 2). Finally, \textbf{P2} can uniformly randomize his remaining attack resource among the remaining 3 monitoring sets that still have untargeted components.

In particular, $k^* = 3$ in this example, and an attack strategy $\sigma^{2*}$ constructed from Lemma~\ref{lemma:cycling} is given as follows:
% In particular, $k^* = 3$ in this example, and the following attack strategy $\sigma^{2*}$ satisfies condition (\ref{NE:attack_prob}) of Theorem \ref{thm:NE}:
\begin{align*}
&\sigma^{2*}_T = \begin{cases}
    \frac{1}{3} & \tm{if } T = T_0   \cup \{e_{1, 3}\} \\
    \frac{1}{3} & \tm{if } T = T_0 \cup \{e_{2, 3}\} \\
    \frac{1}{3} & \tm{if } T = T_0 \cup \{e_{3, 3}\} \\
    0 & \tm{otherwise,}
\end{cases}
\end{align*}
where $T_0 = \{e_{1,1},e_{2,1},e_{3,1},e_{4,1},e_{5,1},e_{1,2},e_{2,2},e_{3,2},$ $e_{4,2}\}$.
We note that $\sigma^{2*}$ satisfies conditions \eqref{NE:attack_prob}.

Since \textbf{P1} has $4>k^*$ sensors, she cycles the positioning of her 3 most accurate sensors among the nodes $v_1$, $v_2$, $v_3$, and deterministically positions her remaining sensor at $v_4$. The construction of such an equilibrium inspection strategy $\sigma^{1*}$ from Lemma~\ref{lemma:cycling} is given as follows:
\begin{align*}
&\sigma^{1*}_s = \begin{cases}
    \frac{1}{3} & \tm{if } s = (v_1,v_2,v_3,v_4) \\
    \frac{1}{3} & \tm{if } s = (v_2,v_3,v_1,v_4) \\
    \frac{1}{3} & \tm{if } s = (v_3,v_1,v_2,v_4) \\
    0 & \tm{otherwise.}
\end{cases}
\end{align*}

The NE $(\sigma^{1*},\sigma^{2*})$ is illustrated in Figure~\ref{fig:sigma2}.
In this example, sensor 1 (in green) has accuracy $\lambda_1 = 0.9$, sensor 2 (in yellow) has accuracy $\lambda_2 = 0.5$, sensor 3 (in orange) has accuracy $\lambda_3 = 0.4$, and sensor 4 (in maroon) has accuracy $\lambda_4 = 0.2$.

\begin{figure}[htbp]
    \centering
    \begin{tikzpicture}
    \node at (4.25, 3) {$b_2 = 10$};
    
    \foreach \y in {0, ..., 4}
    {
        \draw (0, 0.5*\y) rectangle (0.5 , 0.5 + 0.5*\y);
        \foreach \z in {0, ..., 1}
        {
        \draw [color=red, very thick] (0, 0 + 0.5*\z) -- (0.5, 0.5 + 0.5*\z) ;
        \draw [color=red, very thick] (0, 0.5 + 0.5*\z) -- (0.5, 0 + 0.5*\z) ;    
        }
    }
    \node at (0.25 , -0.25) {$v_1$};
    
    \foreach \y in {0, ..., 3}
    {
        \draw (0 + 1, 0.5*\y) rectangle (0.5 + 1, 0.5 + 0.5*\y);
        \foreach \z in {0, ..., 1}
        {
        \draw [color=red, very thick] (0 + 1, 0 + 0.5*\z) -- (0.5 + 1, 0.5 + 0.5*\z) ;
        \draw [color=red, very thick] (0 + 1, 0.5 + 0.5*\z) -- (0.5 + 1, 0 + 0.5*\z) ;  
        }  
    }
    \node at (0.25 + 1 , -0.25) {$v_2$};
    
    \foreach \y in {0, ..., 3}
    {
        \draw (0 + 2, 0.5*\y) rectangle (0.5 + 2, 0.5 + 0.5*\y);
        \foreach \z in {0, ..., 1}
        {
        \draw [color=red, very thick] (0 + 2, 0 + 0.5*\z) -- (0.5 + 2, 0.5 + 0.5*\z) ;
        \draw [color=red, very thick] (0 + 2, 0.5 + 0.5*\z) -- (0.5 + 2, 0 + 0.5*\z) ;  
        }    
    }
    \node at (0.25 + 2, -0.25) {$v_3$};    
    
    \foreach \y in {0, ..., 1}
    {
        \draw (0 + 3, 0.5*\y) rectangle (0.5 + 3, 0.5 + 0.5*\y);
        \foreach \z in {0, ..., 1}
        {
        \draw [color=red, very thick] (0 + 3, 0 + 0.5*\z) -- (0.5 + 3, 0.5 + 0.5*\z) ;
        \draw [color=red, very thick] (0 + 3, 0.5 + 0.5*\z) -- (0.5 + 3, 0 + 0.5*\z) ;  
        }    
    }
    \node at (0.25 + 3, -0.25) {$v_4$};
    
    \foreach \y in {0, ..., 0}
    {
        \draw (0 + 4, 0.5*\y) rectangle (0.5 + 4, 0.5 + 0.5*\y);
        \foreach \z in {0, ..., 0}
        {
        \draw [color=red, very thick] (0 + 4, 0 + 0.5*\z) -- (0.5 + 4, 0.5 + 0.5*\z) ;
        \draw [color=red, very thick] (0 + 4, 0.5 + 0.5*\z) -- (0.5 + 4, 0 + 0.5*\z) ;  
        }  
    }
    \node at (0.25 + 4, -0.25) {$v_5$};    
    
    % \foreach \y in {0, ..., 1}
    % {
    %     \draw (0 + 5, 0.5*\y) rectangle (0.5 + 5, 0.5 + 0.5*\y);
    %     \foreach \z in {0, ..., 1}
    %     {
    %     \draw [color=red, very thick] (0 + 5, 0 + 0.5*\z) -- (0.5 + 5, 0.5 + 0.5*\z) ;
    %     \draw [color=red, very thick] (0 + 5, 0.5 + 0.5*\z) -- (0.5 + 5, 0 + 0.5*\z) ;  
    %     }    
    % }
    % \node at (0.25 + 5, -0.25) {$v_6$};  
    
    % % \node at (7, -0.25) {\uu{  }{ \text{The attacks remaining}}};
    % \draw[decoration={brace, mirror, raise=5pt}, decorate, very thick] (5.75,0) -- node[below=6.5pt] {$\substack{b_2 - \sum_{j=4}^6 |E_{v_j}| \\ \text{remaining attacks}}$ } (8.75, 0);

    \foreach \x in {0, ..., 0}
    {
        \draw [color=red, very thick] (1.05 + 0.5*\x, 2.5 + 0.5 ) -- (1.45 + 0.5*\x, 2.5 + 0.9 ) ;
        \draw [color=red, very thick] (1.05 + 0.5*\x, 2.5 + 0.9) -- (1.45 + 0.5*\x, 2.5 + 0.5) ;
    }
    
    \draw[->, thick] (1.25, 3) -- (0.25, 1.25); 
    \draw[->, thick] (1.25, 3) -- (1.25, 1.25); 
    \draw[->, thick] (1.25, 3) -- (2.25, 1.25); 

    \node at (0.75, 2.6) {\footnotesize $\frac{1}{3}$};
    \node at (1.1, 2.3) {\footnotesize $\frac{1}{3}$};
    \node at (1.75, 2.6) {\footnotesize $\frac{1}{3}$};
    
    % \foreach \y in {0, ..., 5}
    % {
    %     \draw [color=red, very thick] (7, 0 + 0.5*\y) -- (7.4, 0.4 + 0.5*\y) ;
    %     \draw [color=red, very thick] (7, 0.4 + 0.5*\y) -- (7.4, 0 + 0.5*\y) ;
    % }
    % \foreach \y in {0, ..., 4}
    % {
    %     \draw [color=red, very thick] (8, 0 + 0.5*\y) -- (8.4, 0.4 + 0.5*\y) ;
    %     \draw [color=red, very thick] (8, 0.4 + 0.5*\y) -- (8.4, 0 + 0.5*\y) ;
    % }
    
    % \draw [color=red, very thick] (7, 0) -- (7.5, 0.5) ;
    % \draw [color=red, very thick] (7, 0.5) -- (7.5, 0) ;
    
    % \node at (-1, 1) {asdf};
    \foreach \y in {1, ..., 5}
    {
        \node at (-1, 0.23 + 0.5*\y - 0.5) {Layer \y};    
    }
  
    \node at (0.25 , -0.7){\includegraphics[scale=0.1]{images/Sensor_green.png}};
    \node at (0.25 + 1, -0.7){\includegraphics[scale=0.1]{images/Sensor_yellow_2.png}};
    \node at (0.25 + 2, -0.7){\includegraphics[scale=0.1]{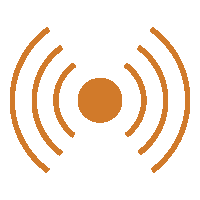}}; 
    \node at (0.25 + 3, -0.7){\includegraphics[scale=0.1]{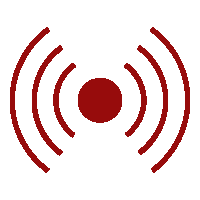}};     
    
    \draw[->, red, thick] (0.25+2+0.4, -0.7) -- (0.25+2+0.5, -0.7) -- (0.25+2+0.5, -0.7) -- (0.25+2+0.5, -1.2) -- (-0.3, -1.2) -- (-0.3, -0.7) -- (-0.08, -0.7);

    \node at (0.25+1, -1.5) {cycle};  
  
\end{tikzpicture}
    \caption{Example of a NE.}
    \label{fig:sigma2}
\end{figure}

In this NE, we observe that the 3 most accurate sensors are randomized so that the detection probability of each node in $\{v_1,v_2,v_3\}$ has an identical detection probability given by $\frac{1}{3}(0.9+0.5+0.4) = 0.6$. We also note that node $v_5$ is never monitored in this NE. The expected number of attacks in each of the monitoring sets $E_{v_1}$, $E_{v_2}$, $E_{v_3}$ is given by $\frac{7}{3}$. Every component in the remaining monitoring sets is deterministically targeted. Thus, the value of the game $\Gamma$, i.e., the expected number of undetected attacks in equilibrium, for this example is given by $10 - 3\times 0.6\times \frac{7}{3} - 0.2\times 2 = 5.4.$
\hfill $\triangle$
\end{example}

Theorem~\ref{thm:NE} demonstrates that there are scenarios where it is beneficial for \textbf{P1} to leave some components completely unmonitored and instead allocate her resources on parts of the network where there will be a larger number of attacks. Such scenarios occur when $k^* < n$, i.e., when the number of attack resources $b_2$ is large enough and the monitoring sets are of heterogeneous sizes.

% Namely, it states that in some networks, if $b_2$ is large enough then there exists a $j\in \mathbb{N}$ such that, in equilibrium, the defender should never place sensors at the nodes $v_{b_1 + j}, \; v_{b_1 + (j+1)}, \dots, v_n$.

% Conversely, when the number of attack resources is small enough, an interesting consequence of Theorem~\ref{thm:NE} is given in the following corollary: \md{This new corollary is not the converse of what has been mentioned in the previous paragraph. This text here needs to change: 

Conversely, when $k^*=n$, which occurs if and only if $b_2 < n \left| E_{v_n} \right|$, \textbf{P1} randomizes her sensors over all the nodes in the network and monitors every component with identical probability. In fact, in such cases, we have the following result:

% \begin{corollary} \label{corollary: low number of attacks}
%     There exists an equilibrium inspection strategy that cycles the placement of sensors over the entire set of nodes $V$ if and only if $b_2 \leq n |E_{v_n}|$. \bm{Is this not too informal?} \bm{Also I don't think this is only obtainable as a result from theorem 3.1; we need other stuff, so it might be confusing to a reader who doesn't see the if and only if part following from the theorem. Should I still just leave the paragraph before this corollary as is?}\md{This is too informal indeed.}
% \end{corollary}

% \begin{corollary}\label{corollary: low number of attacks}
%     If $b_2 \leq n |E_{v_n}|$, then there exists an equilibrium inspection strategy $\sigma^{1*}$ where the detection probabilities of every node under $\sigma^{1*}$ are equal. \bm{better?}
% \end{corollary}

\begin{corollary}
    The set of equilibrium inspection strategies is identical for any number of attack resources satisfying $b_2 < n|E_{v_n}|$.
\end{corollary}

% \bm{Should we include a remark about the following statement?: If the defender does not know exactly how many attack resources the attacker has, but knows that its a small amount $\leq n\left| E_{v_n} \right|$, then it doesn't matter; she will do the same strategy. This is interesting because its like the defender doesn't have to have complete information about the attacker's capabilities.}\md{Yes, this is a good remark}

Hence, if \textbf{P1} does not know the exact number of attack resources \textbf{P2} has at his disposal, but knows that $b_2 < n |E_{v_n}|$, then she can compute an equilibrium inspection strategy by simply assuming that $b_2=1$. 
% This is a consequence of the fact that $k^* = n$ if and only if $b_2 < n|E_{v_n}|$.
% \bm{How about we rewrite the previous two sentences to be:
% ``
% Hence if \textbf{P1} does not know the exact number of attack resources \textbf{P2} has at his disposal, but knows that $b_2 < n |E_{v_n}|$, then she can computer an equilibrium inspection strategy by simply assuming that $b_2 = 1$. This is a consequence of the fact that $k^* = n$ if and only if $b_2 < n |E_{v_n}|$.
% "}

% \bm{Also mention here that the detection probabilities are the same for this and that $k^*$ is $n$ here.}\bm{How about worded like this?}

Next, we investigate conditions under which \textbf{P2} needs to use all of his $b_2$ resources in equilibrium when the monitoring sets are mutually disjoint:

% \begin{proposition}\
%     \begin{itemize}
%         \item[--] If there exists $t\in [k^*]$ such that $\lambda_t < 1$, then for any equilibrium attack strategy $\sigma^{2*}$ and $T\in A_2$, we have $\sigma^{2*}_T > 0$ only if $|T| = b_2$.

%         \item[--] Conversely, if $\lambda_t = 1$ for all $t\in [k^*]$, then for any $b_2 > \sum_{j=k^*+1}^n |E_{v_j}| + k^* |E_{v_{k^*+1}}|$, an equilibrium attack strategy is given by any attack plan $T^*$ of size $\sum_{j=k^*+1}^n |E_{v_j}| + k^* |E_{v_{k^*+1}}|$ that satisfies
%         $$\forall j \in [n], \ |T^* \cap E_{v_j}| = \min\{|E_{v_{j}}|,|E_{v_{k^*+1}}|\}.$$

%     \end{itemize}

% \end{proposition}

% \bm{Is prop 2 a newer version of prop 1??}

\begin{proposition}
    If $b_1 \geq k^*$ and $\lambda_j =1$ for every $j \in [k^*]$, then for any $b_2 > k^* |E_{v_{k^*+1}}| + \sum_{j=k^*+1}^n |E_{v_j}|$, an attack plan $T^*$ of size $k^* |E_{v_{k^*+1}}| + \sum_{j=k^*+1}^n |E_{v_j}|$ that satisfies
        $$\forall j \in [n], \ |T^* \cap E_{v_j}| = \min\{|E_{v_{j}}|,|E_{v_{k^*+1}}|\}$$
        is an equilibrium attack strategy.
        
        Otherwise, for any $b_2 \leq |E|$, any equilibrium attack strategy $\sigma^{2*}$ necessarily randomizes over attack plans $T$ of size exactly $b_2$.

\end{proposition}

This proposition shows that if \textbf{P1} has at least $k^*$ sensors with perfect detection accuracy, then \textbf{P2} does not need to utilize more than $k^* |E_{v_{k^*+1}}| + \sum_{j=k^*+1}^n |E_{v_j}|$ attack resources in equilibrium. Indeed, any additional attack resource would be necessarily allocated to components monitored by perfect sensors, and hence will be detected with probability 1. Therefore, simply targeting $\min\{|E_{v_j}|,|E_{v_{k^*+1}}|\}$ components within each monitoring set $E_{v_j}$ ensures a maximum expected number of undetected attacks in equilibrium.

Finally, the following proposition shows that \textbf{P1} must always use all her sensors in equilibrium:
% \bm{well what if $b_1 > n$, then edge case need to account for. Or should we just state as an assumption that $b_1 \leq n$?}
\begin{proposition}
    For any $b_1 \leq n$, any equilibrium inspection strategy $\sigma^{1*}$ necessarily randomizes over sensor positionings $s \in A_1$ such that $s_k \neq 0$ for all $k \in [b_1]$.
\end{proposition}
From this proposition, we conclude that in any NE, \textbf{P1}'s inspection strategy must randomize over sensor positionings that utilize all her resources when $b_1 \leq n$.
% \bm{when $b_1 \leq n$}

\section{General Case Approximation}\label{sec:General}

\subsection{Solution Approach}

In this section, we leverage our equilibrium results in the case of disjoint monitoring sets to design a heuristic approach for computing an approximate equilibrium inspection strategy in general. In the general case when monitoring sets are not necessarily disjoint, the main challenge lies in determining the subset of nodes that should receive sensors in equilibrium.
As observed in Section~\ref{sec:Disjoint}, \textbf{P2} aims to spread his attacks to maximize the number of undetected attacks. Therefore, \textbf{P1}'s incentive is to position her sensors on nodes that collectively monitor a large number of network components. 
% \bm{Not sure the implication is clear enough}.

One natural candidate set of nodes to receive sensors is given by a \emph{minimum set cover}, i.e., a set of nodes $S \in 2^V$ of minimum size that collectively monitors all network components. Minimum set covers can be obtained by solving the following optimization problem, which can be formulated as an integer program:
$$\min_{S \in 2^V} |S| \text{ subject to } \cup_{v \in S}E_v = E.$$

Although the minimum set cover problem is NP-hard, modern mixed-integer optimization solvers can be used to optimally solve large-scale problem instances \cite{DBLP:journals/corr/DahanSA17}.

To utilize our results in Section~\ref{sec:Disjoint}, we must recreate an instantiation where the monitoring sets are mutually disjoint. To this end, we partition the set of network components by utilizing the monitoring sets of the nodes in a minimum set cover $S = \{v_1^\prime,\dots,v_m^\prime\}\in 2^V$. In Theorem~\ref{thm:NE}, we observed that \textbf{P2} cannot spread his attacks as much in the disjoint case when the monitoring sets are of heterogeneous sizes, thus leading to a lower expected number of undetected attacks. Hence, we partition the set of network components into $m$ subsets by greedily assigning each component to the largest monitoring set containing that component. Specifically, we first determine the monitoring set $E_v$, $v \in S$ of maximum size, suppose it is $E_{v_1^\prime}$, and then remove every component that belongs to $E_v\cap E_{v_1^\prime}$ (for all $v \in S\setminus\{v_1^\prime\}$) from $E_v$. We then repeat this process with the second largest monitoring set, and so on until each network component belongs to exactly one set in the partitioning.

Once this partitioning is obtained, we have an instance with $m$ disjoint monitoring sets. From this, we construct an inspection strategy $\sigma^{1^\prime}$ according to Lemma~\ref{lemma:cycling} that satisfies \eqref{NE:detection_prob} in Theorem \ref{thm:NE}. Since equilibrium inspection strategies are optimal solutions of $(\mathcal{P})$ (see Section~\ref{sec:Problem}), we evaluate the performance of our approximate inspection strategy $\sigma^{1^\prime}$ by computing its objective value in $(\mathcal{P})$, i.e., $\max_{T \in A_2} U(\sigma^{1^\prime},T)$. This determines the worst-case expected number of undetected attacks if \textbf{P1} selects $\sigma^{1^\prime}$ as her inspection strategy.
Since for every attack plan $T \in A_2$, $U(\sigma^{1^\prime},T) = \sum_{e \in T}U(\sigma^{1^\prime},e)$, the largest number of undetected attacks can be efficiently computed by greedily selecting the $b_2$ components with highest probability of undetection under $\sigma^{1^\prime}$.

Our heuristic approach can be summarized as follows:

\begin{algorithm}[htbp]
% \small
    \SetAlgoLined
    \SetKwInOut{Input}{Input}
     \Input{-- Set of nodes $V$\\ 
     -- Set of components $E$\\
     -- Monitoring sets $E_v, \ v \in V$\\
     -- Number of sensors $b_1 \in \mathbb{N}$\\
     -- Number of attack resources $b_2 \in \mathbb{N}$\\
     -- Sensors' accuracies $\lambda_k \hspace{-0.07cm}\in \hspace{-0.07cm}(0,1]$, $k \hspace{-0.07cm}\in \hspace{-0.07cm}[b_1]$}
    \KwResult{-- Inspection strategy $\sigma^{1^\prime}\in\Delta(A_1)$}
    % \md{Should we also output the objective value (worst-case expected number of undetected attacks)?}
    
    Compute a minimum set cover $S \hspace{-0.07cm}= \hspace{-0.07cm}\{v_1^\prime\hspace{-0.01cm},\hspace{-0.01cm}\dots\hspace{-0.01cm},\hspace{-0.01cm}v_m^\prime\}$
    
    Set $E_v^\prime \gets E_v,$ $\forall v \in S$
    
    Set $V^\prime \gets S$
    
    \While{$V^\prime \neq \emptyset$}{
    Select $v^\prime \in \arg\max \{|E_v^\prime|, v \in V^\prime\}$
    
    $E_v^\prime \gets E_v^\prime \setminus (E_{v^\prime}^\prime \cap E_v^\prime)$, $\forall v \in V^\prime\setminus\{v^\prime\}$
    
    $V^\prime \gets V^\prime\setminus\{v^\prime\}$
    }
    
    Order the nodes in $S$ so that $\big| E_{v_1^\prime}^\prime \big|\hspace{-0.08cm}\geq \hspace{-0.08cm}\dotsm \hspace{-0.08cm}\geq \hspace{-0.08cm}\big| E_{v_m^\prime}^\prime \big|$
    
    % $k^* \hspace{-0.1cm}= \hspace{-0.1cm}\min \Bigg\{ k\in [m] \, \Bigg| \, \dfrac{b_2  -  \displaystyle\sum_{\mathclap{j=k+1}}^n |E_{v_j^\prime}^\prime|}{k} \hspace{-0.1cm}\geq \hspace{-0.1cm}|E_{v_{k+1}^\prime}^\prime| \Bigg\}$
    
    $k^* \hspace{-0.1cm}\gets \hspace{-0.1cm}\min \Bigg\{ k\in [m] \, \Bigg|\, \dfrac{b_2  -  \displaystyle\sum_{\mathclap{j=k+1}}^n \big| E_{v_j^\prime}^\prime \big|}{k} \hspace{-0.1cm}\geq \hspace{-0.1cm} \big| E_{v_{k+1}^\prime}^\prime \big| \Bigg\}$    
    
    Construct $\sigma^{1^\prime}$ according to Lemma \ref{lemma:cycling}.
    
    \caption{Heuristic Approach} \label{alg:heuristic}
\end{algorithm}

Next, we implement our heuristic approach on an example network and evaluate the performance of the resulting inspection strategies.

% In the next subsection, we consider an example and compare our approximation $\max_{T\in A_2} U(\sigma^{1*}, T)$ to the value of the full game \bm{or should we say ``the value of the game for the original network"} under the conditions $b_1 = 1, \dots, 10$, with sensor $k$ having detection probability $1 - 0.05 (k-1)$. \md{No, these values should not be introduced here.}
% We find the values of the full game \bm{Same question as previous comment} in these scenarios by implementing column generation.
% We then discuss the advantages of our approach.

% \bm{In general for this section, was it clear what the heuristic was, or was it not entirely clear?}

\subsection{Computational Study}

\begin{figure}[htbp]
    \centering
    \includegraphics[scale=0.4]{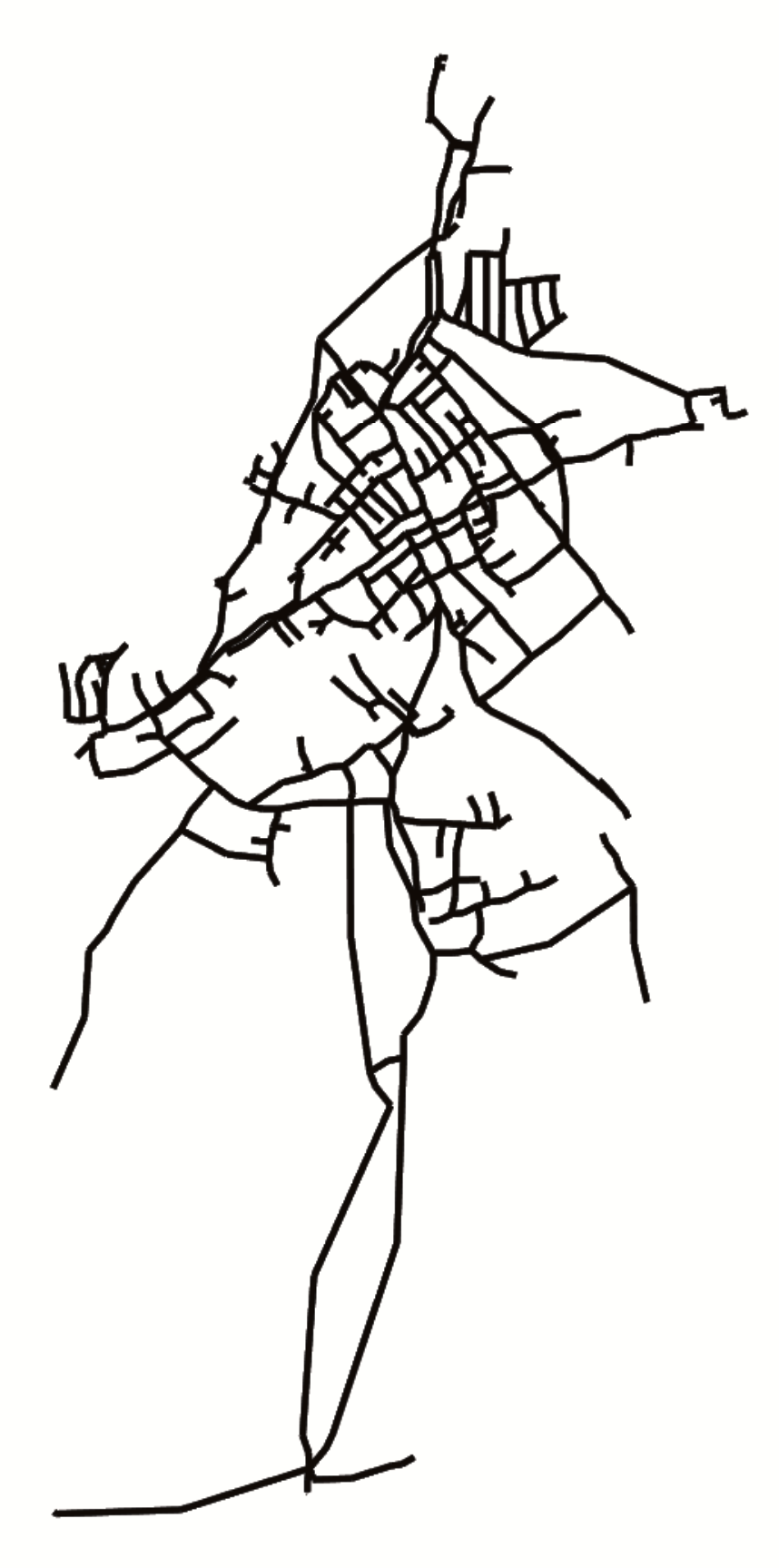}
    \caption{Benchmark Kentucky distribution network.}
    \label{fig:network}
\end{figure}

We consider the benchmark cyber-physical distribution network given in Figure \ref{fig:network}.

% \md{Ne ed to better describe this network}

This real-world network from Kentucky is composed of 420 nodes that can receive sensors, and 492 components that are vulnerable to cyber-physical attacks, which induce disruptions. To detect these attacks, we consider that the defender has access to flow and pressure sensors that can be deployed at access points and shifted from one to another. These sensors can measure signals which can be used to detect the sudden rate of change of pressure or mass flow at different locations of the network. In our study, we compute the monitoring set of each node through simulations using a threshold-based detection model, as proposed in \cite{sensors:autom, deshpande2013optimal}. All network simulations were implemented in Matlab, and all optimization problems were solved using Gurobi on a computer with a 2.3 GHz 8-Core Intel Core i9 processor and 32 GB of RAM.

To evaluate the performance of our heuristic approach we consider 10 game instances where \textbf{P2} has $b_2 = 1$ attack resource and \textbf{P1} has $b_1 \in [10]$ sensors, with sensor $k \in [b_1]$ having accuracy $\lambda_k = 1 - 0.05 (k-1)$. For such instances, $(\mathcal{P})$ only has 494 constraints since $b_2 = 1$. Therefore, equilibrium inspection strategies of $\Gamma$ can be obtained by solving $(\mathcal{P})$ using the column generation algorithm. 

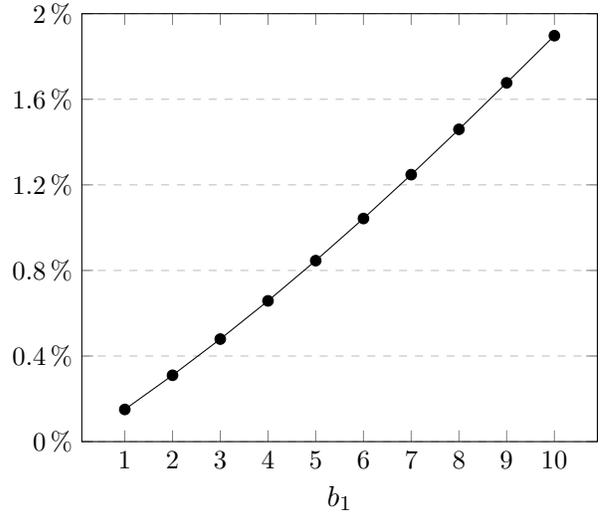
\begin{figure}[htbp]
    \centering
    \pgfplotsset{y tick label style={/pgf/number format/.cd, fixed, precision=4}}   %Found here: https://tex.stackexchange.com/questions/602592/ticks-on-the-y-axis-not-working-as-anticipated
    \begin{tikzpicture}
        \begin{axis}[name=plot, ymin=0, ymax=2, xtick={0, ..., 10}, ytick={0, 0.4, 0.8, 1.2, 1.6, 2}, yticklabel=\pgfmathprintnumber{\tick}\,$\%$, ymajorgrids=true,
            grid style=dashed, xlabel={\large$b_1$}]
            \addplot[black, mark=*]
            table{./data/Opt_Gap.txt};\label{ratio_approximated_value_versus_optimal_value}
        \end{axis}
    \end{tikzpicture}
    
    \caption{Optimality gap of the heuristic solution when $b_2=1$.}
    \label{fig:ratio_of_game_to_optimal_value_plot}
\end{figure}

We now implement our heuristic approach: We solve the minimum set cover problem, and obtain a set of 19 nodes. Next, we greedily partition the set of network components into 19 sets. Finally, we construct an inspection strategy $\sigma^{1^\prime}$ according to Lemma~\ref{lemma:cycling}. The worst-case expected number of  undetected attacks under the inspection strategy $\sigma^{1^\prime}$ is then computed by selecting the $b_2$ components with the highest probability of not being detected under $\sigma^{1^\prime}$. In Figure~\ref{fig:ratio_of_game_to_optimal_value_plot}, we illustrate for $b_1 \in [10]$ the optimality gap achieved by $\sigma^{1^\prime}$, i.e., the relative difference between the worst-case performance of $\sigma^{1^\prime}$ and the value of the game (given by the optimal value of $(\mathcal{P})$).

From Figure~\ref{fig:ratio_of_game_to_optimal_value_plot}, we observe that our heuristic solution achieves a detection performance that is close to the detection performance in equilibrium. However, we note that as the number of sensors increases, the optimality gap associated with our heuristic solution increases. This is due to the fact that when \textbf{P1} has more sensors, she can strategically coordinate their positioning so as to maximize the detection probabilities of the components that are monitored from multiple locations. In contrast, our heuristic approach assigns such components to a single monitoring set to construct an inspection strategy using a disjoint instance.

\begin{figure}[htbp]
    \centering
    \pgfplotsset{scaled y ticks=false}
    \begin{tikzpicture}
        \begin{axis}[name=plot, ymin=0, ymode=log, ymajorgrids=true,
            grid style=dashed, xtick={0, ..., 10}, xlabel={\large$b_1$}]
            \addplot[black, mark=*] table{./data/running_time_optimal_solution.txt};\label{opt_value_running_time}
            \addplot[red, mark=*] table{./data/running_time_approximated_solution.txt};\label{approximated_value_running_time}  
        \end{axis}
        
        \node[anchor=north, fill=white, draw=black, thick] at ($(plot.south) - (0 mm, 11 mm)$)
        {\begin{tabular}{l l}
            Optimal Solution & \ref{opt_value_running_time}\\
            Heuristic Solution & \ref{approximated_value_running_time}
        \end{tabular}
        };
        
    \end{tikzpicture}
    \caption{Running times (in seconds) of the column generation algorithm and the heuristic method.}
    \label{fig:approximated_versus_optimal_running_time_plot}
\end{figure}
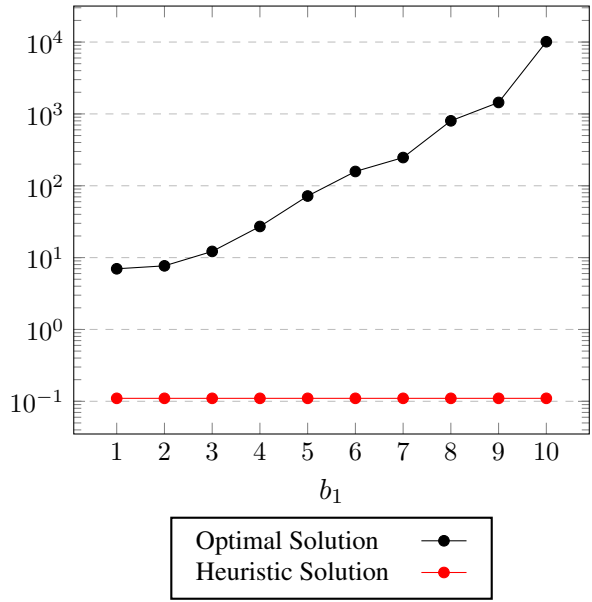

Next, we compare in Figure~\ref{fig:approximated_versus_optimal_running_time_plot} the running times of our heuristic method with the running times of the column generation algorithm for computing equilibrium inspection strategies.

Interestingly, we observe that our heuristic solution is obtained in 0.11 seconds, and this running time is almost identical for any number of sensors. The reason is that most of the running time is spent computing a minimum set cover. As previously mentioned, although this problem is NP-hard, it can be efficiently solved by modern mixed-integer optimization solvers. In contrast, the time required to compute an equilibrium inspection strategy using column generation increases exponentially with the number of sensors $b_1$. This is due to the fact that the number of variables in $(\mathcal{P})$ grows combinatorially with respect to $b_1$. For instance, when $b_1 = 10$, the number of variables in $(\mathcal{P})$ is is approximately $1.54\cdot 10^{26}$ for this network.

Finally, we note that the column generation algorithm for computing equilibrium inspection strategies cannot be used in practice when $b_2 > 1$, as the number of constraints in $(\mathcal{P})$ grows combinatorially with respect to $b_2$. By leveraging the analytical characterization derived in Section~\ref{sec:Disjoint}, our heuristic approach remains scalable for any value of $b_1$ and $b_2$, and can be implemented for large-scale networks, as minimum set covers have been shown to be efficiently solvable for networks containing more than 100,000 nodes and components \cite{DBLP:journals/corr/DahanSA17}.

% as minimum set covers have been shown to be solvable efficiently for large-scale networks

% and can be implemented for large-scale networks, as minimum set covers have been shown to be solvable efficiently \cite{DBLP:journals/corr/DahanSA17}.

% to obtain our approximate solution is much faster than the running time to obtain the exact solution, especially as the number of sensors $b_1$ increases.
% Furthermore, finding the exact solution can be intractable for $b_2 > 1$, while our approximation will have the same running time for any value of $b_2$ \bm{While not \textit{exactly} the same running time (may be up to n more calculations) its the same big-O running time. Should I specify that or is it fine as is?}. \md{big-O running time is exponential due to minimum set cover}
% Finally, the exact solution is obtained using column generation, which does not computationally scale well with respect to the parameters of the problem.
% On the other hand, our approximate solution scales well with respect to the parameters of the problem.
% These facts support our approach of using Theorem \ref{thm:NE} with our minimum set covers heuristic to obtain a NE which gives close approximations for the value of the game when the monitoring sets in our network are not necessarily mutually disjoint.

\section{Conclusion}
\label{sect:pdf}

In this paper, we studied a network inspection game in which a defender allocates sensors with potentially heterogeneous detection capabilities in order to detect multiple attacks caused by a strategic attacker.
In this two-person zero-sum game, the defender
(resp. attacker) seeks to minimize (resp. maximize) the expected number of undetected attacks by selecting a potentially randomized inspection (resp. attack) strategy.
When the monitoring sets are mutually disjoint, we derived an analytical characterization of a class of NE for this game. Additionally, we studied the dependence of these NE on the network topology, sensor accuracies, and the number of resources the attacker has at his disposal. We then leveraged our equilibrium analysis to design a heuristic solution approach for the general case based on minimum set covers. Our computational study on a benchmark cyber-physical distribution network showed that our heuristic approach is computationally tractable and provides inspection strategies with good detection performance. In future work, we aim to refine our heuristic solution approach and provide theoretical performance guarantees.

\bibliographystyle{ieeetr}
\bibliography{bibliography}

\end{document}